\documentclass[fleqn,usenatbib]{mnras}

\usepackage{newtxtext,newtxmath}

\usepackage[T1]{fontenc}
\usepackage{ae,aecompl}

\usepackage{graphicx}	
\usepackage{amsmath}	
\usepackage{amssymb}	

\newcommand{\advspres}{Adv. Space Res.}

\newcommand{\nrph}{Nature Rev. Phys.} 
\newcommand{\nata}{Nature Astron.}

\newcommand{\univ}{Univ.}

\title[A crucial test of the PhDEC cosmological model]
{A crucial test of the phantom closed cosmological model}

\author[S. I. Shirokov et al.]{
S. I. Shirokov$^\text{1}$\thanks{E-mail: arhath.sis@yandex.ru}
and Yu. V. Baryshev,$^\text{2}$
\\
$^\text{1}$SPb Branch of Special Astrophysical Observatory of Russian Academy of Sciences, 65 Pulkovskoye Shosse, St. Petersburg, 196140, Russia\\
$^\text{2}$Saint Petersburg State University,
7/9 Universitetskaya Nab., St. Petersburg, 199034, Russia\\
}

\date{Accepted 2020 September 18. Received 2020 September 18; in original form 2020 July 14}

\pubyear{2020}

\begin{document}

\label{firstpage}
\pagerange{\pageref{firstpage}--\pageref{lastpage}}
\maketitle

\begin{abstract}
We suggest a crucial direct-observational test for measuring distinction between the standard $\Lambda$CDM model and recently proposed  phantom dark energy positive curvature cosmological model. The test is based on the fundamental distance--flux--redshift relation for general Friedmann models. It does not depend on the CMBR data, on the large-scale structure growth models, and also on the value of the Hubble constant $H_0$. 
Our crucial test can be performed by future gamma-ray burst  observations with THESEUS space mission and by using gravitational-wave standard siren observations with modern advanced LIGO--Virgo and also forthcoming LISA  detectors.
\end{abstract}
%



\section{Introduction}

In the last few years there has been growing evidence for a number of ``tensions'' between the derived parameters of the early Universe and measured parameters of the late local Universe. Both the cosmic microwave background radiation (CMBR) data and the local Universe observations have revealed underlying discrepancies that cannot  be ignored.

It comes from comparison of the measured Hubble Constant $H_0$ for the early and late Universe, so-called ``the $H_0$ tension'' \citet{Riess2020,Verde2019,Lin2019}, and  also, from uncertainty in curvature density parameter -- ``the curvature tension'' ~\cite{Handley2019}.


Recent analysis of the combined observational data, using a high-redshift Hubble diagram for Type Ia supernovae, quasars, and gamma-ray bursts,
also discovered  a tension with the flat 
$\Lambda$CDM model
\citep{Lusso2019,Risaliti2019,Demianski2020,Demianski2017a,Demianski2017b}.


Further evidence on modern ``crisis'' in cosmology was found from recent combined analysis of the Planck CMB power spectra
anisotropy and large-scale structure (LSS) data 
~\citep{DiValentino2020a, DiValentino2020b}.
Their results cast doubt on the standard values of basic $\Lambda$CDM parameters for inflation, non-baryonic dark matter, and dark energy. Instead of the flat universe with cosmological constant $\Lambda$,  they suggest to consider the phantom dark energy closed (PhDEC)   cosmological model.

The conclusion of the recent works ~\citet{DiValentino2020a, DiValentino2020b,Handley2019, Riess2020,Verde2019,Lin2019}
is that either $\Lambda$CDM needs to be replaced by a drastically different model, or else there is significant but still undetected systematics. The new  theoretical suggestions call for new observations and stimulate the investigation of alternative theoretical models and solutions.

Here we suggest a crucial cosmological test for measuring distinction between the standard $\Lambda$CDM and proposed PhDEC cosmological models. It does not use the CMBR data, the LSS growth, and the value of the Hubble constant.  General analysis of  CMBR independent tests for alternative cosmological models was considered by ~\cite{Baryshev2012,Shirokov2020}. 
In this letter, we apply this approach to the phantom dark energy positive curvature cosmological model, which now is considered as possible alternative to the standard $\Lambda$CDM model.

\section{Testing the PhDEC cosmological model}

\subsection{Motivation for the independent testing PhDEC model}

Recent discussion on the $\Lambda$CDM ``tensions'' points to consideration of a wider class of cosmological models,
which are based on reanalysis of the CMBR data and N-body simulations of the LSS growth
\citep{Riess2020,Verde2019,Lin2019,DiValentino2020a, DiValentino2020b,Handley2019}. Also, problems in  the galaxy formation theory raise the issue on necessity for consideration a more general initial conditions in the cosmological N-body simulations
\cite{Peebles2020,Benhaiem2019}.

The PhDEC model was suggested in~\citet{DiValentino2020a, DiValentino2020b} as a ``new standard'' Universe model. They performed reanalysis of the Planck-2018 temperature and polarization CMBR angular power spectra, the cosmic shear measurements, and  the baryon acoustic oscillation data.    
They also emphasized existing uncertainties in determination of the Hubble constant value $H_0$~\citep{Riess2020,Verde2019,Lin2019}.


In view of the tension between different approaches to CMBR and LSS data analyses, here we suggest a robust observational test of the reality of the phantom dark energy closed (positive curvature) cosmological model, which is independent from CMBR data, LSS data and the determination of the local value of the Hubble constant.
 
As the independent test of the PhDEC model, we consider the high-redshift Hubble diagram, which is based on the directly observed flux--distance--redshift relation for a sample of standard candles
~\citep{Amati2018,Demianski2020}. 
Such test belongs to the class of classical crucial cosmological tests, which related to the basis of the cosmological models
\citep{Baryshev2012}.
It allows us to perform robust testing Friedmann--Lemaitre--Robertson--Walker (FLRW) primary quantities such as curvature constant, density and equation of state (EoS) for dark energy and dark matter
~\citep{Shirokov2020,Demianski2020}.

This crucial test can be performed by future gamma-ray burst (GRB) observations with THESEUS space mission~\citep{Amati2018},
and by using gravitational-wave standard siren observations with LIGO--Virgo and LISA advanced detectors~\citep{Schutz1986,Holz2005b,Abbott2017}.

\subsection{Basic formulas of the PhDEC model}

Friedmann--Lemaitre--Robertson--Walker expanding space cosmological model is a direct consequence of  the general relativity and cosmological principle of the homogeneous and isotropic distribution of matter. 

The general form of the Friedmann equation is given by the formula
\begin{equation}
    H^2 - \frac{8\pi G}{3}\rho =
    -\frac{kc^2}{S^2} \, , \qquad \text{or} \qquad
    1 - \Omega = -\Omega_k \, ,
    \label{Friedmann-eq}
\end{equation}
where $k = (-1,\,0,\,+1)$ is the curvature constant for open, flat, or closed space, respectively, $S$ is the scale factor, 
$H=(\text{d}S/\text{d}t)/S$ is the Hubble parameter,
$\rho = \Sigma_i\,\rho_i$ is the total density of all cosmological homogeneous non-interacting fluids $\rho_i$. 

The total matter
density parameter is defined as 
$\Omega = \Sigma_i\,\Omega_i$, where
$\Omega_i=\rho_i/\rho_\text{c}$ with the critical density 
$\rho_\text{c} = 3H^2/8\pi G$.
The curvature density parameter is defined as
\begin{equation}
    \Omega_k  = kc^2/S^2H^2 = 
    (\Omega -1) \,.
    \label{curvature density-eq}
\end{equation}
We use definition of $\Omega_k$ which has the same sign as the curvature constant 
$k$, i.e. positive curvature space ($k=+1$)
corresponds to the positive curvature density parameter $\Omega_k > 0$. 
\footnote{
Note that in number of papers it is also used another definition
$\Omega'_k = -\Omega_k$,
so positive curvature space ($k=+1$)
corresponds to the negative 
curvature density parameter $\Omega'_k < 0$.}

The $w$CDM model is defined as the FLRW model that contains  two cosmological non-interacting  fluids, having EoS of the cold matter $p=0$, and the quintessence (dark energy) $p=w\rho c^2$ (with $w<0$). Thus, the  normalised Hubble parameter $h(z)=H(z)/H_0$ 
is given by Eq.~\ref{Friedmann-eq} 
in the form
\begin{equation}
    h(z)=\sqrt{\Omega^0_\text{m}(1+z)^3 +
    \Omega^0_\text{DE}(1+z)^\text{3(1+w)} -
    \Omega^0_k(1+z)^2}\,,
    \label{Hubble-$w$CDM}
\end{equation}
where $\Omega^0_i$ is the corresponding density parameter at present epoch,
and $w$ is the dark energy EoS parameter. 
 
For $w=-1$ one have $p=-\rho c^2$ and constant cosmological vacuum density
$\Omega_\Lambda =\Omega^0_\text{DE}= \text{const}$. In this case, $w$CDM model is called the $\Lambda$CDM model. For curvature constant $k=0$ and  
($\Omega_\Lambda =0.7,\,\Omega_\text{m} =0.3 $)
one have the standard  $\Lambda$CDM model.

If dark energy parameter $w < 0$, the model is called quintessence $w$CDM. For $w<-1$, one have so-called   ``phantom'' $w$CDM model. In the case of curvature constant $k=+1$, the model is called phantom dark energy closed FLRW model.

\subsection{Relative luminosity distance}

Our test of the PhDEC model is based on the measurement of the luminosity distance $d_\text{L}$ in the Friedmann models, which is given by the following expression
\begin{equation}
  d_\text{L}(z)=
  \frac{c}{H_0}
  \frac{(1+z)}{\sqrt{|\Omega^0_k|}}I_k
  \left(  \sqrt{|\Omega^0_k|}
  \int_0^z \frac{\text{d}z'}{h(z')} \right)\,,
    \label{lum-dist-L}
\end{equation}
where $k=(-1,0,+1)$ is the curvature constant, $\Omega^0_k$ is the curvature density parameter, $I_k(x)=\sinh(x)$ for $\Omega_k<0$, and $I_k(x)=x$ for $\Omega_k=0$, and $I_k(x)=\sin(x)$ for $\Omega_k>0$. 

Let us consider the relative luminosity distance 
$ D_\text{rel}$, i.e.
the ratio of the  luminosity distance of a considered cosmological model to the luminosity distance of the fixed standard flat $\Lambda$CDM model
($\Omega^0_k=0,\,\Omega^0_\Lambda =0.7,\,\Omega^0_m =0.3,\,w=-1$).
So, we get the equation
\begin{equation}
    D_\text{rel}(z) =
   \frac{d_\text{L}(z)}{d_\text{$\Lambda$CDM}(z)}
   = F(z;\,\Omega^0_k ,\, \Omega^0_\text{DE},\, w ) \,,
    \label{rel-d-L}
\end{equation}
which depends only on three parameters  
($\,\Omega^0_k ,\, \Omega^0_\text{DE},$ and $w$)
of the considered cosmological  model.
Thus, the relative luminosity distance $ D_\text{rel}$ does not depend on the Hubble constant $H_0$.  The matter density parameter 
$\Omega_\text{m}$ does not enter to the Eq.~(\ref{rel-d-L}) and is determined by the Friedmann relation  Eq.~(\ref{Friedmann-eq}) as 
\begin{equation}
    \Omega_\text{m} = \Omega_k- 
    \Omega_\text{DE} + 1 \,.
    \label{omega-m}
\end{equation}

The relative luminosity distance modulus is given by relation $5\log  D_\text{rel}(z)$\,, i.e. it is equal to the difference between a considered model of luminosity distance modulus and the standard $\Lambda$CDM  luminosity distance modulus
\begin{equation}
    \Delta \mu_\text{L}(z;\,\Omega^0_k ,\, 
    \Omega^0_\text{DE},\, w ) =
    \,5\log\,F(z;\,\Omega^0_k ,\, \Omega^0_\text{DE},\, w )\,. 
    \label{delta-mu-L}
\end{equation}

Eq.~(\ref{delta-mu-L}) depends on the three parameters of a studied FLRW model
$\,\Omega^0_k ,\, \Omega^0_\text{DE}$, and $w$ (for fixed standard $\Lambda$CDM model) and it can be used as a test of the phantom dark energy closed cosmological model
independently from the~\citet{DiValentino2020a, DiValentino2020b} analysis.

\section{Results of calculations}

To estimate the predicted relative luminosity distance modulus $\Delta \mu_\text{L}(z)$ for a number of phantom dark energy FLRW models having closed, flat, and open geometry we consider following parameters: ~$\Omega^0_k=(-0.1,\,0.0,\,+0.1)$, 
~$\Omega^0_\text{DE}=(0.3,\,0.5,\,0.7,\,0.9)$,
and ~$\,w=(-1,\,-2)$.

\begin{figure*}
    \centering
    \includegraphics[height=0.49\textwidth,angle=-90]{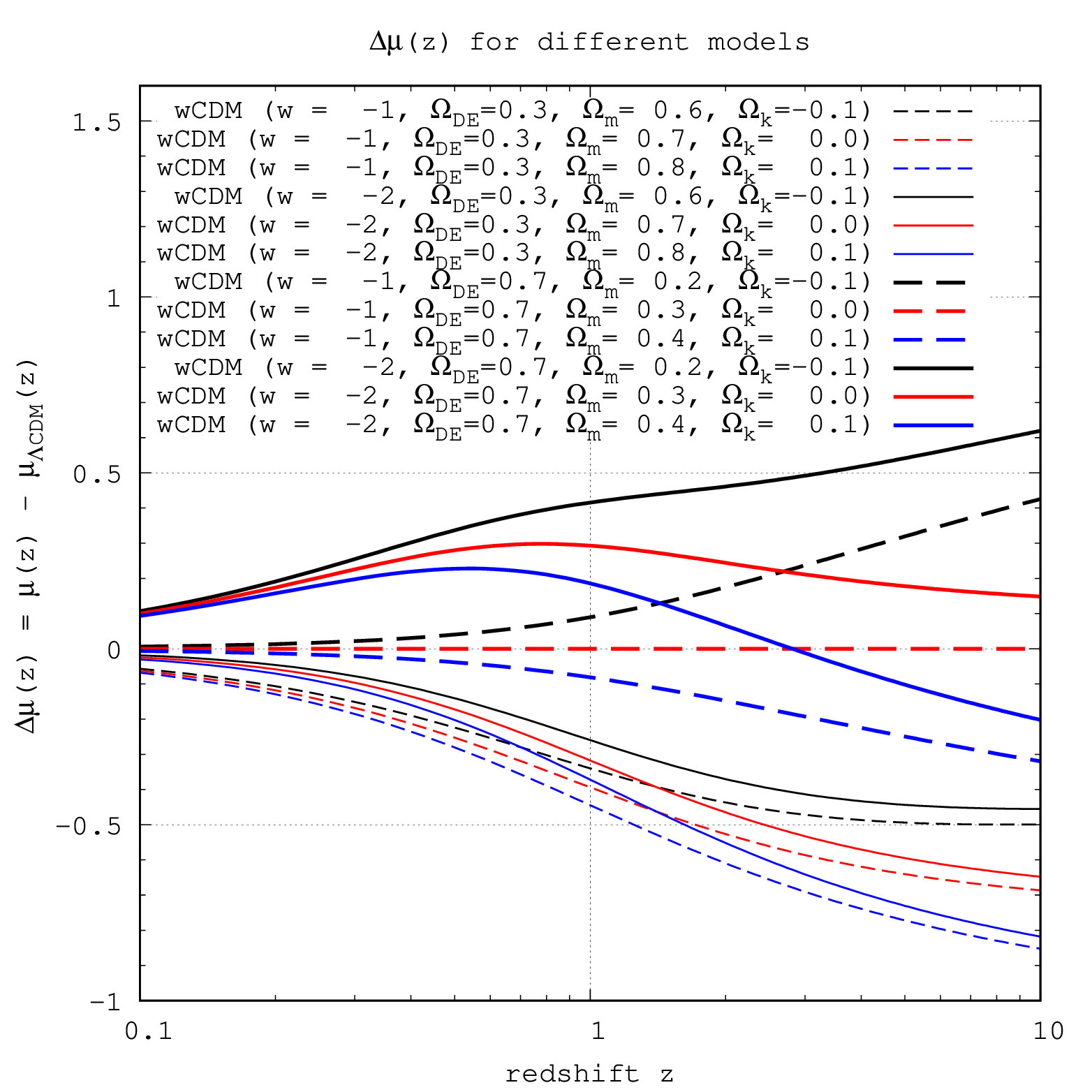}
    \hfill
    \includegraphics[height=0.49\textwidth,angle=-90]{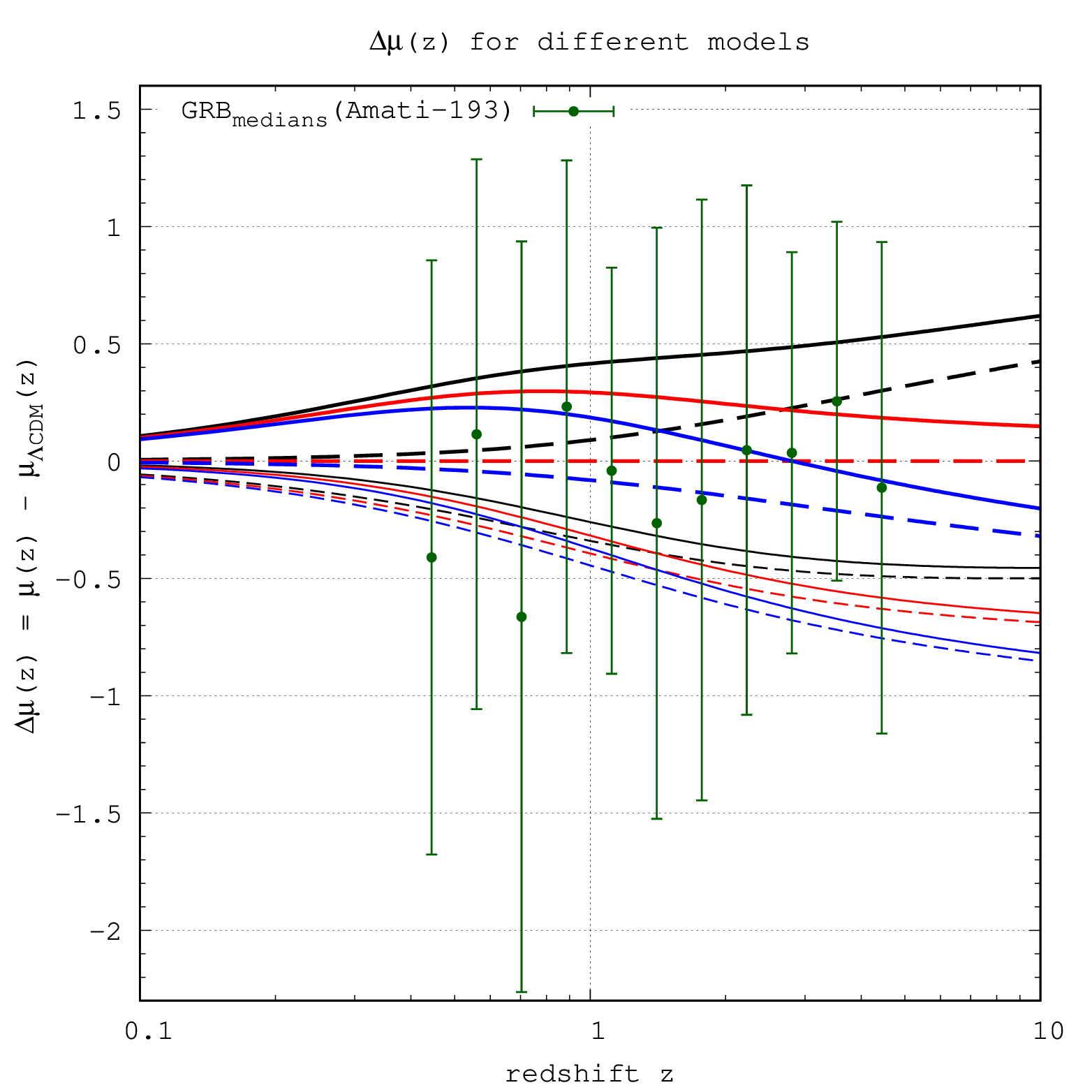}
    \caption{
    \emph{Left:} The relative distance modulus $\Delta \mu_\text{L}(z)$ verse redshift is shown. Predicted quantitative difference between standard $\Lambda$CDM and phantom $w$CDM models for the case
    $\Omega^0_\text{DE}=(0.3,\,0.7),\,w=(-1,\,-2)$ and $k=(-1,\,0,\,+1)$. 
    \emph{Right:} The one with medians of the 193-GRB sample~\citep{Amati2019} for $\Delta \log{z}=0.1$.
    }
    \label{fig:HD_borders}
\end{figure*}

Fig.~\ref{fig:HD_borders} shows predicted quantitative difference between standard $\Lambda$CDM and phantom $w$CDM models for the case 
$\Omega^0_\text{DE}=(0.3,\,0.7),\,w=(-1,\,-2)$ in redshift logarithmic scale. 
On the right panel, the available median data on  high-redshift long GRBs from the~\citet{Amati2019} 193-GRB sample are also shown.

Table~\ref{tab:mu} contains values of the relative luminosity distance modulus  $\Delta \mu_\text{L}(z)$  for phantom dark energy EoS with $w = -2$.

\begin{table*}\centering
\begin{tabular}{c|cccccccccccc} \hline \hline

\multicolumn{13}{c}{Phantom dark energy EoS parameter $w = -2$} \\
\hline
$\Omega_k$ & $-0.1$ & $0.0$ & $+0.1$ & $-0.1$ & $0.0$ & $+0.1$ & $-0.1$ & $0.0$ & $+0.1$ & $-0.1$ & $0.0$ & $+0.1$  \\

$\Omega_\text{DE}$ & 0.3 & 0.3 & 0.3 & 0.5 & 0.5 & 0.5 & 0.7 & 0.7 & 0.7 & 0.9 & 0.9 & 0.9 \\

$\Omega_\text{m}$ & 0.6 & 0.7 & 0.8 & 0.4 & 0.5 & 0.6 & 0.2 & 0.3 & 0.4 & 0.0 & 0.1 & 0.2 \\

\hline

 $z=0.10$ & -0.02 & -0.02 & -0.03 &  0.04 &  0.04 &  0.03 &  0.11 &  0.10 &  0.09 &  0.18 &  0.17 &  0.16 \\
 0.20 & -0.05 & -0.06 & -0.07 &  0.06 &  0.05 &  0.03 &  0.19 &  0.17 &  0.16 &  0.35 &  0.33 &  0.31 \\
 0.30 & -0.08 & -0.10 & -0.12 &  0.07 &  0.05 &  0.03 &  0.25 &  0.23 &  0.20 &  0.50 &  0.46 &  0.42 \\
 0.50 & -0.14 & -0.17 & -0.20 &  0.06 &  0.02 & -0.02 &  0.34 &  0.28 &  0.23 &  0.78 &  0.67 &  0.59 \\
 0.60 & -0.17 & -0.21 & -0.24 &  0.05 &  0.01 & -0.04 &  0.36 &  0.29 &  0.23 &  0.89 &  0.75 &  0.64 \\
 0.80 & -0.22 & -0.27 & -0.31 &  0.03 & -0.03 & -0.09 &  0.40 &  0.30 &  0.21 &  1.10 &  0.87 &  0.70 \\
 1.00 & -0.26 & -0.32 & -0.37 &  0.01 & -0.07 & -0.14 &  0.42 &  0.29 &  0.19 &  1.27 &  0.95 &  0.72 \\
 2.00 & -0.37 & -0.47 & -0.55 & -0.05 & -0.18 & -0.29 &  0.46 &  0.24 &  0.07 &  1.86 &  1.08 &  0.68 \\
 3.00 & -0.41 & -0.53 & -0.64 & -0.07 & -0.23 & -0.38 &  0.49 &  0.21 & -0.01 &  2.25 &  1.12 &  0.60 \\
 4.00 & -0.43 & -0.57 & -0.69 & -0.07 & -0.26 & -0.43 &  0.52 &  0.19 & -0.06 &  2.57 &  1.13 &  0.54 \\
 5.00 & -0.44 & -0.59 & -0.73 & -0.07 & -0.28 & -0.47 &  0.54 &  0.18 & -0.10 &  2.84 &  1.13 &  0.49 \\
 6.00 & -0.45 & -0.61 & -0.76 & -0.07 & -0.30 & -0.49 &  0.56 &  0.17 & -0.13 &  3.08 &  1.14 &  0.45 \\
 8.00 & -0.45 & -0.63 & -0.79 & -0.06 & -0.32 & -0.53 &  0.59 &  0.16 & -0.17 &  3.48 &  1.14 &  0.40 \\
10.00 & -0.46 & -0.65 & -0.82 & -0.06 & -0.33 & -0.56 &  0.62 &  0.15 & -0.20 &  3.82 &  1.15 &  0.35 \\

\hline

\end{tabular}                     
    \caption{
    Predicted quantitative difference $\Delta \mu_\text{L}(z)$ between standard $\Lambda$-CDM and phantom $w$-CDM models for the case     $\Omega^0_\text{DE}=(0.3,\,0.5,\,0.7,\,0.9),\,w=-2$ and $k=(-1,\,0,\,+1)$.
    }\label{tab:mu}
\end{table*}

\section{Discussion and conclusions}

The high-redshift relative luminosity distance modulus
$\Delta \mu_\text{L}(z)$ ~~
(Eq.~\ref{delta-mu-L} and 
Fig.~\ref{fig:HD_borders}) can be used as a crucial test of the phantom dark energy closed universe model (PhDEC) that recently was suggested as a possible alternative to the standard flat $\Lambda$CDM model~\citep{DiValentino2020a, DiValentino2020b}.

An advantage of $\Delta \mu_\text{L}(z)$ test is that it does not depend on the  CMBR anisotropy analysis and on theory of the density fluctuation growth. Our suggested test is based on directly observed flux--distance--redshift relation for general FLRW model.
The $\Delta \mu_\text{L}(z)$-test depends only on three parameters 
~($\,\Omega^0_k ,\, \Omega^0_\text{DE}$, and $w$) ~of the tested FLRW models, which determine the specific shape of the corresponding curves.

Our calculations of the $\Delta \mu_\text{L}(z)$  for the number of FLRW $w$CDM models give quantitative estimation of needed measurement accuracy for detecting the distinction between the standard flat $\Lambda$CDM and PhDEC models. 
In our test one must use the whole shape of the function (Eq.~\ref{delta-mu-L}) within studied redshift interval 
($0.1 <z< 10$).
According to our calculation presented at Fig.~\ref{fig:HD_borders}, the accuracy of distance modulus measurements $\Delta \mu \sim 0.1$ at $z\sim 1$ is enough for distinction between considered models.
Future observations of high-redshift GRBs and gravitational waves (GW) will bring such possibility for redshift interval $0.1<z<10$.

For small redshifts  ($z<0.1$) the Hubble Diagram is used as a measure of the local Hubble constant value $H_0$ because of the linear relation 
$z = (H_0 /\,c)\, r$ is fulfilled  for all models. So according to
Eq.~(\ref{delta-mu-L})
the $\Delta \mu_\text{L}(z)$-test does not depend on the value of the $H_0$.

To performing the $\Delta \mu_\text{L}(z)$-test, one need observations of the objects having standard luminosity  in the high-redshift interval $0.1<z<10$. Possible candidates for such objects are the long GRBs. 
Fig.~\ref{fig:HD_borders} (right)
shows how the existing sample of 193 long GRBs \citep{Amati2019}  with calibrated luminosity by the Amati relation fits the models (with error bars for median intervals of $\Delta \log{z}=0.1$). 
High luminosity quasars also can be used as a standard candles for the Hubble diagram analysis
~\citep{Lusso2019,Risaliti2019}.
Future THESEUS GRB observations will reduce the error bars and can distinguish between standard $\Lambda$CDM and PhDEC.

Perspective candidates for $\Delta \mu_\text{L}(z)$-test are the gravitational wave standard sirens, which are 
the GW analog of the optical ``standard candles'' 
\citep{Schutz1986, Holz2005b, Abbott2017}.
We should emphasized that important obstacle for measuring distances to the standard luminosity objects is the necessity  of accounting the gravitational lensing bias 
\citep{Broadhurst2020}. However, forthcoming  surveys of the weak and strong gravitational lensing, like EUCLID, will help to overcome this problem
\citep{CervantesCota2020}.

\section*{Acknowledgements}
We thank referee for useful comments and suggestions. 
The work was performed as part of the government contract of the SAO RAS approved by the Ministry of Science and Higher Education of the Russian Federation.  

\section*{Data availability}

The data underlying this article will be shared on reasonable request to the corresponding author.

\bibliography{references}

\bsp	
\label{lastpage}
\end{document}